\documentclass[aps,prl,twocolumn,groupedaddress,showpacs]{revtex4}

\usepackage{graphicx}
\usepackage{amsmath}

\newcommand{\hata}{\hat{a}}
\newcommand{\hatad}{\hat{a}^{\dag}}
\newcommand{\hatb}{\hat{b}}
\newcommand{\hatbd}{\hat{b}^{\dag}}
\newcommand{\ket}[1]{| #1 \rangle}
\newcommand{\meanI}[1]{\langle #1 \rangle}

\begin{document}

\title{Violation of the Cauchy-Schwarz Inequality in the Macroscopic Regime}

\author{A. M. Marino}
\author{V. Boyer}
\author{P. D. Lett}

\affiliation{Joint Quantum Institute, National Institute of
Standards and Technology,\\ and University of Maryland,
Gaithersburg, Maryland 20899, USA}

\date{\today}

\begin{abstract}
We have observed a violation of the Cauchy-Schwarz inequality in the
macroscopic regime by more than 8 standard deviations.  The
violation has been obtained while filtering out only the low
frequency noise of the quantum-correlated beams that results from
the technical noise of the laser used to generate them. We use
bright intensity-difference squeezed beams produced by four-wave
mixing as the source of the correlated fields. We also demonstrate
that squeezing does not necessarily imply a violation of the
Cauchy-Schwarz inequality.
\end{abstract}


\maketitle


The comparison between the predictions of quantum and classical
theories has been a subject of study since the development of
quantum mechanics. To that end, a number of different classical
inequalities have been developed that provide an experimental
discrimination between these theories~\cite{Loudon80,Reid86}.
Experiments showing a violation of these classical inequalities have
verified quantum theory. However, to date, most of these experiments
have been carried out in the regime in which single particles are
detected one at a time. It is thus interesting to study whether or
not the quantum signature given by these tests is still present in
the limit in which the system under study becomes macroscopic.

Among the inequalities that offer a test between quantum and
classical theories is the Cauchy-Schwarz inequality
(CSI)~\cite{Loudon80,Reid86}. The first observation of a violation
of this inequality was obtained by Clauser using an atomic
two-photon cascade system~\cite{Clauser74}. More recently, large
violations using four-wave mixing have been
obtained~\cite{Kolchin06,Thompson06}, still in the photon-counting
regime. For bright fields the natural approach for analyzing their
quantum nature is through noise measurements.  In this case the
boundary between quantum and classical is taken to be the noise of a
coherent state, or standard quantum limit (SQL), such that having a
field with less noise than the SQL (squeezed light) is considered
non-classical. However, the presence of squeezing does not provide a
direct discrimination between quantum and classical theories since
the SQL is a result of quantum theory~\cite{Reid86}.

The possibility of using a macroscopic quantum state to violate the
CSI has been previously
analyzed~\cite{McNeil83,Gerry95a,An02,Olsen03}. To date, however,
only a few experiments have probed this macroscopic regime. Recently
anti-bunching of a small number of photons was observed in the
continuous-variable regime~\cite{Grosse07}. In addition, a frequency
analysis has been used to infer a violation of the CSI over limited
frequency ranges~\cite{Li00}.

In this Letter we present the first observation, to our knowledge,
of a direct violation of the two-beam Cauchy-Schwarz inequality in
the limit of a macroscopic quantum state. We show that the
quantum-correlated fluctuations between two different modes of the
electromagnetic field are responsible for the violation of the CSI.
In addition to having a bright coherent carrier, we work in the high
gain regime in which the mean number of spontaneous correlated
photons within the inverse of the bandwidth (correlation time) of
the process is much larger than one. Thus photon counting is not an
option and continuous variable detection schemes need to be used.

The CSI for the degree of second-order coherence, $g^{(2)}$, for two
distinct fields, $a$ and $b$, is of the form~\cite{Loudon}
\begin{equation}
    [g^{(2)}_{ab}(\tau)]^{2}\leq g^{(2)}_{aa}(0)g^{(2)}_{bb}(0),
    \label{cs}
\end{equation}
where $g^{(2)}$ is the normalized intensity correlation function.
This inequality indicates that for a classical system the
cross-correlation between two fields, $g^{(2)}_{ab}$, cannot be
larger than the geometric mean of the zero-time auto-correlations,
$g^{(2)}_{aa}$ and $g^{(2)}_{bb}$.  According to quantum theory,
however, it is possible to violate this inequality.  In this case
the correlation function is defined in terms of normally ordered
operators
\begin{equation}
    g^{(2)}_{ab}(\tau)=\frac{\langle\hatad(t)\hatbd(t+\tau)\hatb(t+\tau)\hata(t)\rangle}
    {\langle\hatad(t)\hata(t)\rangle\langle\hatbd(t)\hatb(t)\rangle},
    \label{g2}
\end{equation}
where $\hat{a}$ and $\hat{b}$ are the photon annihilation operators
for the two fields.  A violation of the CSI indicates the presence
of non-classical correlations between the fields.

Most of the experiments to date have been done in the
photon-counting regime, in which the separation between photon pairs
is much larger than the correlation time between photons.  This
makes it possible to obtain a large cross-correlation while the zero
time auto-correlation functions are in principle equal to 2, giving
as a result a large violation of the CSI~\cite{Loudon}. In contrast,
in the large gain regime all of the correlation functions tend to
the same value (they are equal or larger than one), making it harder
to observe a violation of the CSI.

\begin{figure}[h]
\centering
\includegraphics{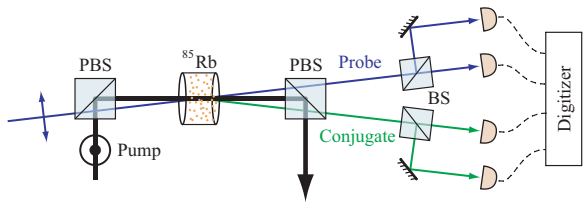}
\caption{(color online). Experimental setup.  A 4WM process is used
to generate quantum-correlated bright beams. PBS~=~polarizing beam
splitter, BS~=~50/50 beam splitter.\label{Setup}}
\end{figure}
We use a seeded four-wave mixing (4WM) process in a double-$\Lambda$
system in rubidium vapor, as described
in~\cite{McCormick07,McCormick07a}, as our source of bright
correlated beams. Four-wave mixing is a parametric process, such
that the initial and final states of the atomic system are the same.
This leads to the emission of probe and conjugate photons in pairs
and thus to intensity correlations between the two fields which are
stronger than any correlations possible between classical optical
fields.

The configuration and experimental parameters for the 4WM are the
same as the ones described in Ref.~\cite{McCormick07a}. A single
Ti:Sapphire laser and an acousto-optic modulator are used to
generate a bright pump and a weak probe which are resonant with a
two-photon Raman transition between the $F = 2$ and $F = 3$
electronic ground states of $^{85}$Rb.  The pump laser is tuned
800~MHz to the blue of the D1 line at 795~nm while the probe is
downshifted in frequency by 3~GHz. The two beams are then mixed at a
small angle in a pure $^{85}$Rb vapor cell, as shown in
Fig.~\ref{Setup}. In our double-$\Lambda$ configuration, the 4WM
converts two photons from the pump into one probe photon and one
conjugate photon (upshifted by 3~GHz with respect to the pump). We
have measured up to 8~dB of intensity-difference squeezing at 1~MHz
with this method.

After the vapor cell we separate the probe and conjugate from the
pump beam with a polarizer with $\approx 10^{5} : 1$ extinction
ratio for the pump..  We then use beamsplitters to split the probe
and conjugate, each into two beams of equal power, and detect the
resulting four beams with separate photodiodes, as shown in
Fig.~\ref{Setup}. This setup directly measures the normally ordered
correlation function defined in Eq.~(\ref{g2}), as described in
Ref.~\cite{Reid86}. After each photodiode a bias-T is used to
separate the DC part of the photocurrent, which is recorded and then
used to normalize the correlation functions.  The rest of the signal
is amplified, digitized with a resolution of 9 bits, and recorded on
a computer. The amplified time traces are sampled at a rate of
1~GS/s and 500 sets of traces, each with 10,000 points, are
recorded. This setup allows us to simultaneously obtain all the
information needed to calculate the correlation functions and the
noise power spectra of the different beams.

The bright correlated beams that are obtained from the seeded 4WM
process consist of a large coherent part plus quantum-correlated
fluctuations. The large coherent part makes the $g^{(2)}$ functions
tend to 1, the value for a coherent state, as its intensity
increases. It is thus useful to separate the correlation functions
into contributions for the coherent part of the field and the
fluctuations, that is $g^{(2)}_{ab}=1+\epsilon_{ab}$. Since the
quantum correlations between the fields are in the fluctuations, we
can rewrite the CSI in terms of the fluctuation terms such that it
takes the form
\begin{equation}
    \epsilon_{ab}\leq \frac{\epsilon_{aa}+\epsilon_{bb}}{2},
\end{equation}
where we have kept only terms to first order in $\epsilon$.  We
define a violation factor
\begin{equation}
    V\equiv \frac{\epsilon_{aa}+\epsilon_{bb}}{2\epsilon_{ab}}
\end{equation}
such that $V < 1$ indicates a violation of the CSI.

In the ideal case, the 4WM process can be described by the
two-photon squeeze operator $\hat{S}_{ab} = \exp(s\hat{a}\hat{b} -
s\hat{a}^{\dag}\hat{b}^{\dag})$, where $s$ is the squeezing
parameter ($s>0$). The bright quantum-correlated beams are obtained
by applying this operator to an input coherent state,
$\ket{\alpha}$, for the probe and the vacuum for the conjugate. In
the limit in which the number of photons in the probe seed is much
larger than one ($|\alpha|\gg1$) $V$ takes the form
\begin{equation}
    V=1-\frac{1}{2G},
    \label{V}
\end{equation}
where $G=\cosh ^{2}s$ is the gain of the process and we have taken
the single frequency approximation for each beam. For an ideal
seeded 4WM process $V$ is always less than one, so that a violation
of the CSI should always be obtained. In general, however, the
presence of squeezing does not guaranty a violation of the CSI. As
Eq.~(\ref{V}) shows, the amount of violation is inversely
proportional to the gain. This is in contrast to the amount of
intensity-difference squeezing that is expected from a seeded 4WM
process, for which the noise scales as $1/(2G-1)$. Thus, a large
amount of squeezing does not imply a large violation of the CSI, as
has been pointed out in Ref.~\cite{Carmichael00}.

\begin{figure}[ht]
\centering
\includegraphics{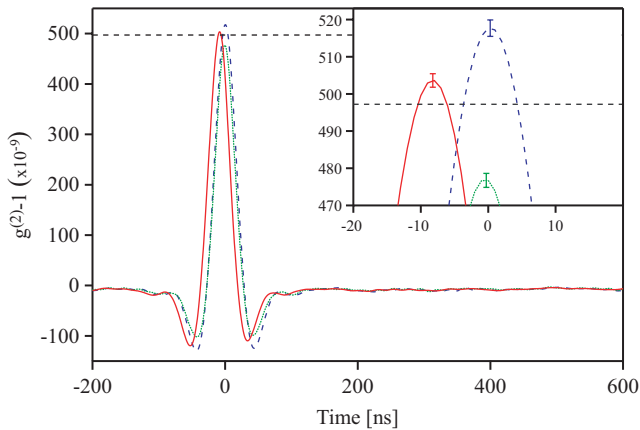}
\caption{(color online). Correlation functions of the fluctuations
for the probe (dotted), conjugate (dashed), and cross $g^{(2)}$
(solid). The inset shows an expanded view of the peaks of the
correlation functions. The horizontal dashed line shows the mean
value of the zero time auto-correlation functions for the probe and
the conjugate. The indicated uncertainties are discussed in the
text.\label{g2s}}
\end{figure}
A typical set of correlation functions that shows a violation of the
CSI is shown in Fig.~\ref{g2s}.  Here the horizontal dashed line
indicates the mean value of the zero-time auto-correlation functions
of the fluctuations for the probe and the conjugate, such that a
cross-correlation larger than this level indicates a violation of
the CSI. A violation of the CSI can clearly be seen in the inset of
Fig.~\ref{g2s}. In obtaining these correlation functions we have
only filtered out the low frequency technical noise below 500~kHz.
The bandwidth of the detection system ($>$ 40~MHz) is larger than
the bandwidth of the quantum correlations.

The uncertainties indicated in Fig.~\ref{g2s} are obtained by
directly calculating the correlation functions and obtaining the
standard deviation over the 500 sets of traces. These uncertainties
are not statistically independent since the probe and conjugate
contain classical fluctuations that are strongly correlated as a
result of slow intensity fluctuations of the pump and probe seed
beams between data sets. This leads to a violation of the CSI that
is more significant than what can be inferred from the inset. An
accurate measure of the uncertainty of $V$ is obtained by
calculating $V$ for each set of traces and using these results to
derive the standard deviation of $V$ over the 500 sets. For the
results shown in Fig.~\ref{g2s} the gain of the process is around 10
and $V=0.987\pm1.4\times10^{-3}$, giving a violation of the CSI by
more than 8 standard deviations.

The cross-correlation function shows a delay in the arrival time
between probe and conjugate fluctuations; for the case shown in
Fig.~\ref{g2s} the delay is around 8~ns.  The delay results from the
combination of 4WM in the double-$\Lambda$ system and propagation
through the vapor cell~\cite{vanderWal03,Boyer07}. An important
property of the double-$\Lambda$ system is that the relative delay
between probe and conjugate for fluctuations of different
frequencies is almost fixed. Such a fixed delay only causes the
cross-correlation to be shifted in time and will not have an effect
on $V$. In contrast, any large spread in the delay between different
frequencies (dispersion) would make the cross-correlation peak wider
and reduce its maximum value, degrading the amount of violation. The
dips on the correlation functions are due to an offset of the
carrier frequency with respect to the gain peak of the process.
These effects will be examined in detail elsewhere.

One of the difficulties in obtaining a violation of the CSI is that
any source of excess uncorrelated noise will decrease the violation.
In order to see why this is the case, we need to consider the noise
power spectra of the different beams. We can rewrite the CSI in
terms of the noise power spectra for the probe ($S_{p}$), conjugate
($S_{c}$), and intensity-difference ($S_{\rm{diff}}$) such that
\begin{align}
    &\int{d\Omega \left( \frac{S_{\rm{diff}}(\Omega)}{\meanI{\hat{n}_p}+\meanI{\hat{n}_c}}-1 \right)}\nonumber\\
    &\geq\frac{\meanI{\hat{n}_p}-\meanI{\hat{n}_c}}{\meanI{\hat{n}_p}+\meanI{\hat{n}_c}}
    \int{d\Omega \left[\left( \frac{S_{p}(\Omega)}{\meanI{\hat{n}_p}}-1
\right) -\left( \frac{S_{c}(\Omega)}{\meanI{\hat{n}_c}}-1
\right)\right]}.
    \label{freqnoise}
\end{align}
The terms in parenthesis represent the excess noise (or noise
reduction) with respect to the corresponding SQL. For the ideal
seeded 4WM process the normalized noise power spectra for the probe
and the conjugate are equal, so that the term in square brackets is
zero, and $\meanI{\hat{n}_p}>\meanI{\hat{n}_c}$. The presence of
squeezing in the intensity difference can make the integral on the
left hand side negative, leading to a violation of the CSI.  Excess
noise can have an impact on the violation in two different ways. The
presence of excess uncorrelated noise on either beam can lead the
intensity-difference noise to go above the SQL for some frequency
ranges such that the integral on the left hand side can become
positive. In addition, excess noise on the conjugate can make the
right hand side of the inequality negative enough (given that
$\meanI{\hat{n}_p}>\meanI{\hat{n}_c}$) so that even if squeezing is
present a violation might not be obtained.

\begin{figure}[ht]
\centering
\includegraphics{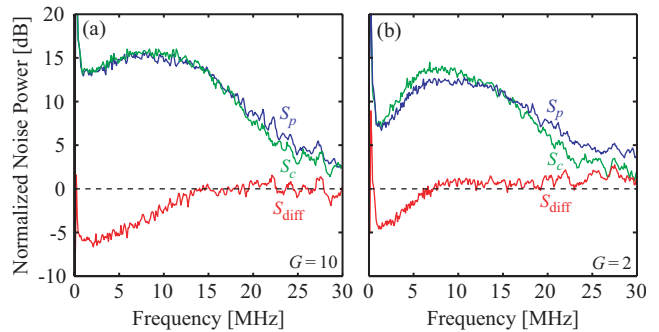}
\caption{(color online). Normalized noise power spectra for the
probe ($S_{p}$), conjugate ($S_{c}$), and intensity-difference
($S_{\rm{diff}}$) for a gain of (a) 10 and (b) 2. All the spectra
are normalized to their respective SQL, represented by the dashed
line.
\label{Spect}}
\end{figure}
For the results shown in Fig.~\ref{g2s}, the corresponding
normalized noise power spectra  are shown in Fig.~\ref{Spect}(a).
All the noise power spectra are calculated by taking the FFT of the
time traces and averaging over the 500 sets. The SQL for the probe
and conjugates is calculated by taking the difference of the
corresponding photocurrents while the one for the
intensity-difference noise is given by the sum of the SQLs for the
probe and conjugate.  As is expected for a 4WM process both the
probe and the conjugate have excess noise with respect to the SQL
and their spectra are almost the same.  The measured
intensity-difference squeezing has a bandwidth of 15~MHz, consistent
with the gain bandwidth of the 4WM process~\cite{Boyer07}, with a
maximum squeezing of 6~dB. For this case the system acts almost as
an ideal 4WM medium which makes it possible to observe a violation
of the CSI. The amount of squeezing that is measured is limited by a
total detection efficiency, including optical path transmission and
photodiode efficiencies, of $(80\pm3)\%$. We have verified that
$g^{(2)}$ is not affected by loss so that any source of loss will
not have an impact on the violation of the CSI.

When the gain of the process is reduced to 2, we find a situation in
which the noise power spectra of the probe and the conjugate are
noticeably different, as shown in Fig.~\ref{Spect}(b). This
difference in noise leads to a reduction of the intensity-difference
squeezing bandwidth from 15~MHz to 7~MHz and a small amount of
excess noise at higher frequencies.  For this particular case we
find that the small amount of excess noise is enough to prevent a
violation of the CSI, such that $V=1.075\pm3.3\times10^{-3}$, even
though there is more than 4~dB of squeezing at low frequencies.

The relative delay between the probe and the conjugate (8~ns for
$G=10$ and 13~ns for $G=2$) has been compensated when calculating
the intensity-difference noise power spectra shown in
Fig.~\ref{Spect}. This makes it possible to see the real squeezing
bandwidth that results from the 4WM process in Fig.~\ref{Spect}(a).
While the relative delay has no effect on the violation of the CSI,
it introduces a frequency dependent phase shift such that the
intensity-difference noise power spectrum oscillates between
intensity-difference and intensity-sum noise
levels~\cite{Machida89}.

\begin{figure}[ht]
\centering
\includegraphics{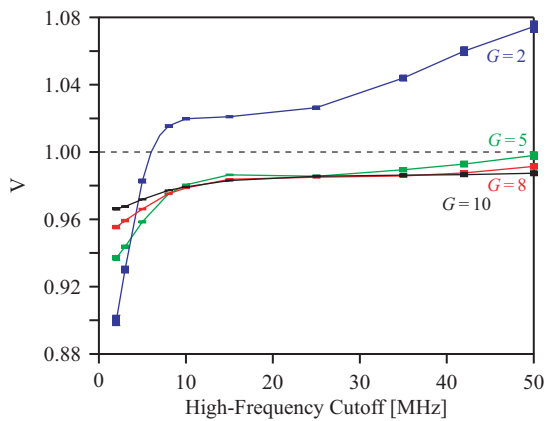}
\caption{(color online). Effect of frequency filtering on the
violation of the CSI. Violation parameter ($V$) as a function of
high-frequency cutoff for $G=2$, $G=5$, $G=8$, and $G=10$. $V<1$
indicates a violation of the CSI. The size of the squares represent
the statistical uncertainties. \label{Vfreq}}
\end{figure}
The effect of the excess noise can be further analyzed by filtering
out the high frequencies, where most of the uncorrelated excess
noise is present.  The filtering is done on the digitized traces by
applying a 10$^{\rm{th}}$ order Butterworth bandpass filter with a
low-frequency cutoff of 500~kHz that filters out the technical noise
of the laser and a variable high-frequency cutoff. We have done this
analysis for a number of different gains, as shown in
Fig.~\ref{Vfreq}. The gain is changed by modifying the temperature
of the cell and thus the atomic number density.

If we look at the lowest high-frequency cutoff points in
Fig.~\ref{Vfreq}, we see that the violation follows the trend given
by Eq.~(\ref{V}) for an ideal 4WM process, that is, the violation
gets better with smaller gains. However, once we increase the
high-frequency cutoff, $V$ starts to degrade, with lower gains
degrading faster. Increasing the high-frequency cutoff takes into
account higher frequencies of the noise power spectrum that
correspond to different regions of the gain profile.  This leads to
competition with other processes, such as Raman gain on the
conjugate, that add excess noise. Except for the case $G=2$ a
violation of the CSI is obtained for the different gains shown in
Fig.~\ref{Vfreq} when only the low frequency technical noise of the
laser is filtered. Even for the case in which the system contains
excess uncorrelated noise, a violation of the CSI can be recovered
with enough filtering, as shown for the case of $G=2$. This
approaches a spectral analysis of the noise, as is regularly done
when measuring bright beams.

If we compare the case of $G=2$ in Fig.~\ref{Spect} and
Fig.~\ref{Vfreq} we find that the violation is lost at a
high-frequency cutoff around 6~MHz while the squeezing is present
over a larger frequency range than the filtering bandwidth used to
calculate V, up to around 7~MHz, once the relative delay between
probe and conjugate has been compensated. This gives a region in
which squeezing is present but not a violation of the CSI. The
amount of excess noise on the conjugate is enough to destroy the
violation but not the squeezing.

In conclusion, we have observed a violation of the CSI in the
macroscopic regime. The necessary information to observe the
violation is contained in the quantum-correlated fluctuations of the
field. We have shown that the presence of excess uncorrelated noise
can prevent the observation of a violation of the CSI. The ability
to obtain a violation of the CSI shows that the 4WM process used
here provides a low-noise source of quantum correlated bright beams
over a large frequency range. Finally, we have shown that the
presence of squeezing does not necessarily imply a violation of the
CSI.


\end{document}